\documentstyle[aps,prl,multicol,]{revtex}
\input {epsf.sty}
\begin{document}   
%\draft
\title{Antiferromagnetism in the Vortex Cores of YBa$_2$Cu$_3$O$_{7-\delta}$}
\author{V. F. Mitrovi{\'c}, E. E. Sigmund,  W. P. Halperin}
\address{Department of Physics and Astronomy,\\
   Northwestern University, Evanston, Illinois 60208}
\author{A. P. Reyes, P. Kuhns, W. G. Moulton}
\address{National High Magnetic Field Laboratory
   Tallahassee, Florida 32310}
%\date{Received 15 April 2000}
\date{Version \today}
\maketitle
\begin{abstract} We report spatially resolved nuclear magnetic resonance measurements on a high temperature superconductor
that indicate the presence of correlated antiferromagnetic fluctuations in the vortex core.  The nuclear
spin-lattice relaxation rate,
$1/^{17}T_1$, of planar
$^{17}O$, in  near-optimally doped  YB$_2$Cu$_3$O$_{7-\delta}$ (YBCO), was measured.  Outside of the core,
$(^{17}T_1T)^{-1}$ is independent of temperature consistent with theoretical predictions for a {\it d}-wave superconductor. 
In the vortex core 
$(^{17}T_1T)^{-1}$  increases with decreasing temperature following an antiferromagnetic Curie-Weiss law.
\end{abstract}
%\narrowtext
\vspace{-11pt}
\begin{multicols}{2}

\def\ie{\emph{i.e.} }
\def\etal{{\it et al.} }
\def\T1{$T_1^{-1}$}
\def\Tw{$T_2^{-1}$}
\def\TT1{($T_1T)^{-1}$}
\def\delinv{$\left( \delta (T_1T)^{-1} \right )^{-1}$}
\def\delT{$\left( \delta (T_1T)^{-1} \right )$}
\def\hi{$H_{int}$}
\def\h0{$H_{0}$}
\def\vq {{\bf q}}
\def\Cu{$^{63}$Cu }
\def\Ox{$^{17}$O }
\def\Brl{Brillouin }
\def\vk {{\bf k}}
\def\Dk {\Delta_{{\bf k}}}
\def\DDk {\Delta^{\ast}_{-\vec{k}}}

High temperature superconductors are commonly viewed as doped Mott
insulators for which the parent compound is antiferromagnetic.  For these
materials competition between magnetism and superconductivity is
unambiguously evident in the phase diagram. Varying the composition by
increasing the oxygen content increases the electronic carrier density,
suppressing the N{\'e}el phase and producing superconductivity.  Even for
those compositions that exhibit superconductivity, the presence of
antiferromagnetic (AF) fluctuations  is apparent in the normal state from
inelastic neutron scattering\cite{dai01}, and copper NMR
relaxation\cite{takigawa89,timusk99,millis90}.   Experiments have shown
that antiferromagnetic fluctuations coexist with superconductivity even
in optimally doped materials (highest
$T_c$)\cite{dai01}. There have been predictions that magnetism can appear, possibly
in a novel, spatially-inhomogeneous form, coexisting with vortices\cite{zhang97,arovas97,demler01,hu01,Herbut02,tesanovic02}.  Using
spatially-resolved, high-field, NMR experiments  we have found  evidence for antiferromagnetism in the vortex cores of YBCO, lending 
support for these ideas.   

In the mixed state, magnetic flux penetrates the sample in the form of quantized vortices, each with a core of radius the size
of the superconducting coherence length, $\xi_o = 16$ \AA.  According to theoretical models
antiferromagnetism can appear in the vortex core, a region where superconductivity is suppressed.   Zhang
\cite{zhang97,hu01} has developed a theory that integrates antiferromagnetism and {\it d}-wave superconductivity based on  SO(5)
symmetry, predicting that the superconducting vortex can have an antiferromagnetic core. Arovas \etal
\cite{arovas97} extended this work to consider the possible coexistence of
  superconducting vortices with antiferromagnetism as a function of doping. Demler
\etal \cite{demler01} have looked in the far field region of the vortex finding that superconducting and coupled
superconducting-spin density wave (SDW) phases can appear in the presence of vortices and that there are oscillations in charge density in
good agreement with STM experiment\cite{hoffman02}.   Charge and spin density wave
structures near, and inside the vortex have been explored theoretically by Zhu {\it et al.}\cite{zhu02}.  They also find results
consistent with experiment\cite{hoffman02}. It has been shown by Herbut\cite{Herbut02} that the {\it d}-wave superconductor is
unstable to the formation of SDW owing to phase disordering in the core of a vortex. These theories have a common feature;  
antiferromagnetism can be associated with vortex cores.

Using elastic neutron scattering Vaknin \etal \cite{vaknin00} found   evidence of   static AF order in the mixed state of YBCO. 
They estimated  that the upper limit of the average magnetic moment is
$\mu \le 0.004\mu_B$ per flux line in a layer.  Katano \etal \cite{katano00} found enhanced static AF correlations in the
vortex state of  La$_{2-x}$Sr$_x$CuO$_4$ (x = 0.12).   Neutron inelastic scattering experiments by Lake \etal
\cite{lake01} on the optimally doped La$_{2-x}$Sr$_x$CuO$_4$ (x = 0.163) show that the imaginary part of the magnetic susceptibility,
$\chi ''$, at low energy below the spin-gap\cite{dai01} and at an incommensurate AF wave vector, is strongly enhanced in a magnetic field,
$H_0 = 7.5$ T at temperatures below
$T = 10$ K.  They interpreted their results as   evidence that the vortex cores are nearly ordered antiferromagnets which polarize  the
intervening medium. These measurements are close to NMR relaxation experiments since both are sensitive to the
same component of the susceptibility. However they do not have spatial resolution.  This is possible with scanning tunneling
microscopy where Hoffman {\it et al.}\cite{hoffman02} have found checkerboard extended states that they associate with vortices although the
method is not specifically sensitive to magnetism.  All of these experiments consistently indicate unusual structure near the vortex core.

Following theoretical suggestions,\cite{takigawa99,wortis00} NMR measurements by Curro {\it et al.}\cite{curro00}, Mitrovi{\'c} {\it
et al.}\cite{mitrovic01} and Kakuyanagi {\it et al.}\cite{kakuyanagi01} as well as
$\mu$SR  lineshape measurements of Miller {\it et al.}\cite{miller01} show that it is possible to spatially resolve
different regions of the vortex lattice by analyzing the internal field distribution of the corresponding resonance spectrum.
However, resolution of the vortex core region is best achieved at relatively high applied fields
since the fraction of the spectrum inside the core grows with increasing field.  For a square vortex lattice this fraction is
$H_0\pi\xi_0^2/\phi_0$, where $\phi_0$ is the flux quantum. In a field of 42 T vortices are
$\sim 86$ \AA $\,$  apart and vortex cores occupy $\sim 17 \%$ of the total sample.  

In the present work we report on the temperature dependence of spatially-resolved
measurements of the nuclear spin-lattice relaxation rate, $1/T_1$, in the mixed state.  This relaxation requires an electronic spin flip and
consequently is particularly sensitive to magnetic fluctuations. Previously we demonstrated\cite{mitrovic01} that we could resolve
the NMR rate in the vortex core at high magnetic field, above 13 T.  Applying this method we have now measured the temperature
dependence of
$(^{17}T_1T)^{-1}$ and extended our range of field. We find that $(^{17}T_1T)^{-1}$ is independent of temperature  outside the vortex cores,
while in the  vortex core region  $(^{17}T_1T)^{-1}$ is enhanced, increasing with decreasing temperature following a Curie-Weiss (CW) law,
reminiscent of the normal state behavior of the copper resonance, $(^{63}T_1T)^{-1}$. We
associate this enhanced temperature  dependence of $(^{17}T_1T)^{-1}$ in the vortex core region with correlated antiferromagnetic
spin fluctuations such as are observed in the normal state.

\begin{figure}[h]
%%%%%%%%%%%%%%%%%%%   F I G U R E  1 %%%%%%%%%%%%%%%%%%%%
\centerline{\epsfxsize0.93\hsize\epsffile{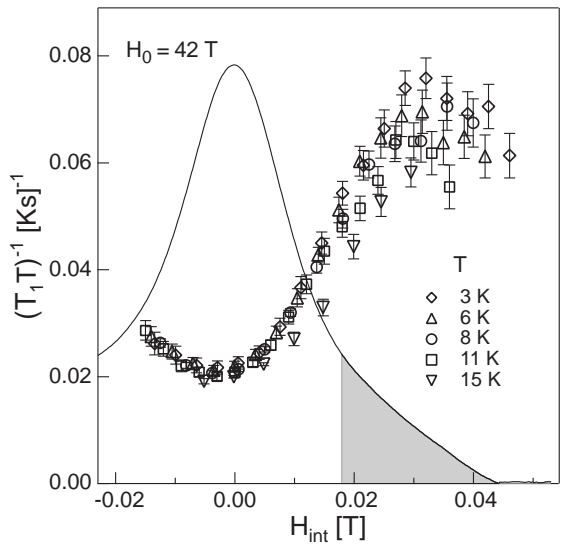}}
%%%%%%%%%%%%%%%%%%%%%%%%%%%%%%%%%%%%%%%%%%%%%%%%%%%%%%%%
\begin{minipage}{0.93\hsize}
\caption[]{\label{TT1TemDep42T}\small Spin-lattice relaxation rate of planar
$^{17}$O divided by temperature  as a function of internal magnetic field.  The (-1/2 to -3/2) satellite spectrum shown is
measured at  37 T, the same as at 42 T, and the shaded region corresponds to the fraction of the spectrum  occupied by vortex
cores at 42 T.  }
\end{minipage}
\end{figure}
\noindent

Our sample is a near-optimally doped, $\sim 60\%$
$^{17}$O-enriched, YBa$_2$Cu$_3$O$_{7-\delta}$ powder sample, aligned with the crystal $\hat c$-axis parallel to
the applied magnetic field. Low-field magnetization data show a sharp  transition at
$T_c(0)=92.5\,\mbox{K}$. Our measurements were made at temperatures from 3 to 25 K and magnetic fields from 6 to 
 42 T.   We have independently determined from spin-spin and spin-lattice relaxation that the transition region in high fields near 30
T occurs at 80 K in our samples. The  (-1/2 to -3/2)
$^{17}$O(2,3) quadrupolar-split NMR satellite was used exclusively since it exhibits a sharp high-field edge in the normal state that
broadens owing to the distribution of local fields from vortices\cite{mitrovic01}.  The internal field variable in Fig. 1, is defined by
$H_{\scriptsize int}={\omega 
\mathord{\left/ {\vphantom {\omega  {{}^{17}\gamma }}}
\right. \kern-\nulldelimiterspace} {{}^{17}\gamma }}-{\it H_0}$ where
$^{17}\gamma$ is the gyromagnetic ratio for oxygen and the spectrometer frequency,
$\omega$, is set to resonance  at the peak of the spectrum,
$H_{int} = 0$.  This position corresponds to oxygen nuclei at the saddle-point of the field distribution, a spatial location
half-way between neighboring vortices.  Increasing
$H_{int}$ corresponds to spatial positions that approach the vortex core.  We obtained precise spectra using  a field
sweep technique\cite{foot} and we measured the relaxation rate  using progressive saturation\cite{mitrovic00}. There is some
overlap between the quadrupolar-split transitions.  Their effect on the spectrum and on relaxation can be
subtracted\cite{mitrovic01}.  In Fig. 2 we report relaxation data for $H_{int} \leq 0.032$ T where  the small correction for the
(-3/2 to -5/2) transition is less than 5\%.  However, the subtracted spectrum in the region near
$H_{int}=0.04$ T in Fig. 1., should be considered to be qualitative. In the present work the main purpose of high
applied field is to increase sensitivity to vortex cores since the corresponding portion of the NMR spectrum grows
proportionately. This is shown for 42 T as the shaded region of the
spectrum in Fig. 1. 

\begin{figure}[h]
%%%%%%%%%%%%%%%%%%%   F I G U R E   2  %%%%%%%%%%%%%%%%%%%%
\centerline{\epsfxsize1.00\hsize\epsffile{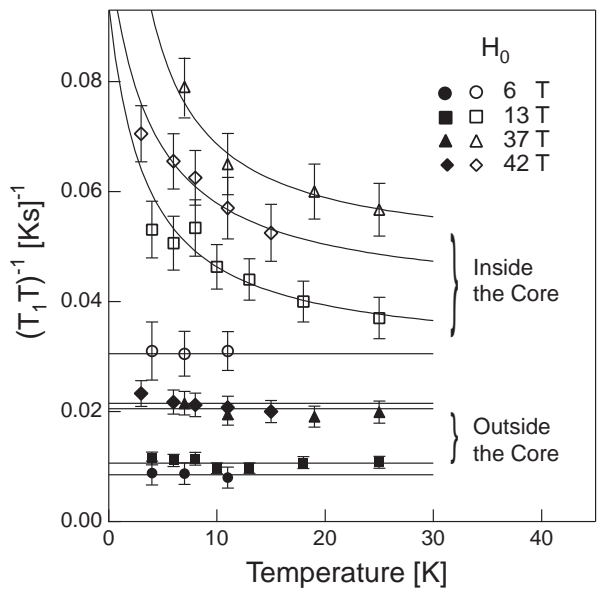}}
%%%%%%%%%%%%%%%%%%%%%%%%%%%%%%%%%%%%%%%%%%%%%%%%%%%%%%%%
\begin{minipage}{0.93\hsize}
\caption[]{\label{TdepVCSP}\small
    Planar $^{17}$O spin-lattice relaxation  extracted from the saddle-point  (full symbols) and the vortex core regions of
the spectrum (open symbols)  versus temperature in 6, 13, 37, and 42 T applied fields. The solid lines are fits to the data
as explained in the text.
 }
\end{minipage}
\end{figure}
\noindent

The spin-lattice relaxation rate at 42 T, shown in Fig. 1, increases as a function of the internal field, \ie on approaching the
vortex cores.  This is attributable\cite{mitrovic01} to a Doppler shift of the quasiparticle excitation energies, a consequence of
the supercurrent momentum that increases toward the vortex core. Here we will concentrate our discussion on  the temperature
dependence of the rate. Outside of the core region we observed that
\TT1 is independent of temperature and at higher internal fields,
in the vortex core, we observe a very different behavior.

In order to contrast the temperature dependence inside and outside the vortex cores in Fig. 1, we show \TT1 in Fig. 2
for  just these two regions of the spectrum in four magnetic fields, 6, 13, 37, and 42 T. 
 The rates outside the cores  were determined by evaluating an average in a narrow  region of \mbox{$\pm 0.002$ T}
around \hi$\,=0$.  Inside the cores our data are averaged over an interval of $H_{int}$ from the
point where \TT1 is greatest to a point that is \mbox{$ 0.01$ T} less than the peak position which occurs for $H_{int} \leq 0.032$.  
 The clear distinction between the temperature dependences of the rate inside and outside
the vortex cores holds for \h0 $\geq$ 13 T.  If the applied field is too low then the sensitivity of our NMR to the vortex core
region is decreased and the enhancement in the temperature dependence of \TT1 is not discernible, or possibly does not
exist,  as might be the case for the 6 T data in Fig. 2.

\begin{figure}[h]
%%%%%%%%%%%%%%%%%%%   F I G U R E  3 %%%%%%%%%%%%%%%%%%%%
\centerline{\epsfxsize0.95\hsize\epsffile{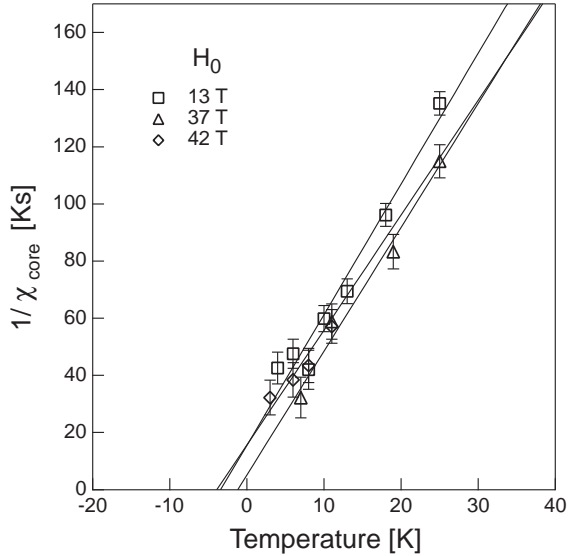}}
%%%%%%%%%%%%%%%%%%%%%%%%%%%%%%%%%%%%%%%%%%%%%%%%%%%%%%%%
\vspace{8pt}
\begin{minipage}{0.93\hsize}
\caption[]{\label{FigChiInv}\small 
The inverse of $\chi_{core}$ versus temperature, defined in the text. The solid lines are fits to the  Curie-Weiss temperature
dependence.}
\end{minipage}
\end{figure}
\noindent

We first discuss \TT1 outside  the vortex core region and give a very simple interpretation of our observations. 
The rate can be viewed as an average of the product over all possible initial and  final quasiparticle density of
states\cite{mitrovic01,wortis00,mitrovicthesis}.
 At low temperature the quasiparticle   excitations outside of the vortex cores   come  exclusively from the four
nodal regions.   Since the density of states depends linearly on energy near the nodes of a {\it d}-wave superconductor, the rate
can be expressed as the product of the initial and
 final quasiparticle excitation energies. The excitation energies are determined by temperature  and, in a
magnetic field, by two other variables:    applied field through the Zeeman effect, and superflow momentum, ${\bf p}_s$
through the Doppler shift\cite{mitrovic01}.     The latter are temperature independent and so, at sufficiently low temperatures, $T
< 20$ K the product of initial and final excitation energies is non-zero at the Fermi energy and temperature independent. It follows
that \TT1 is temperature independent \cite{takigawa99,mitrovic01,mitrovicthesis} as we observe.  Furthermore, the
increase of the rate with applied field shown in Fig. 2, noted previously by Mitrovi{\'c} {\it et al.}\cite{mitrovic01}, is a
consequence of node-to-node quasiparticle scattering indicating the presence of  AF fluctuations.

Inside the core region the temperature dependence of $(T_1T)^{-1}$ in Fig. 2 is quite different from outside
the cores. 
Here we find that \TT1 increases with decreasing temperature.  
This temperature dependence resembles that of the relaxation at the planar copper site in the normal state, which is dominated 
  by correlated AF fluctuations \cite{takigawa89,timusk99,millis90}, and for which the AF correlation length increases with
decreasing temperature. Thus, we suggest that the temperature dependence of $(^{17}T_1T)^{-1}$ in the vortex core is
also dominated  by correlated AF fluctuations in a similar way. Furthermore,   Moriya and Ueda's theory \cite{moriya00} shows that
the temperature dependence  of the imaginary part of the dynamic spin susceptibility, and thus \TT1, for  
 2D antiferromagnetic spin fluctuations of an itinerant electronic system   follows a Curie-Weiss law. This behavior is
central to the phenomenological model of Millis {\it et al.}\cite{millis90} that successfully describes the normal state where
the key parameter is the AF correlation length. Consequently, we proceed with the following analysis of the temperature dependence of
\TT1.  

In the vortex core we find that \TT1 
consists of two parts, a temperature
independent contribution,
$R_0$, and a temperature dependent part, with the form of a Curie-Weiss law, $\chi_{core}$, giving \TT1 $= R_0 + \chi_{core} =
R_0 + C/(T-\theta_{CW})$. In Fig. 3. we plot the inverse of
$\chi_{core}$ as a function of temperature where the lines in the figure are linear fits.
The inverse of the slope of these lines and their zero temperature intercept have a phenomenological interpretation as a Curie constant, $C$,
and Curie-Weiss temperature, $\theta_{CW}$.  Within the accuracy of these fits the Curie-Weiss temperature is negative lying in the
range -1 to -4 K for magnetic fields from 13 to 42 T and the Curie constant is {\it field independent}; consequently,  magnetic
fields above 13 T do not affect the low energy AF fluctuations.  The  dependence of antiferromagnetic fluctuations on doping of a 
2D antiferromagnetic has been shown theoretically to have a quantum critical point\cite{chubukov94} that is observed in normal state 
measurements
$(^{63}T_1)^{-1}$ in LSCO\cite{Imai93}. Similarily,  the enhanced temperature dependence of $(^{17}T_1T)^{-1}$, moving toward the
vortex core, can be interpreted as an approach to a quantum critical point where the amplitude of the
superconducting order parameter plays the role of doping.  It follows from the
theory that a  negative
$\theta_{CW}$ indicates a  spin-gap in the vortex core, while a positive value of $\theta_{CW}$ would
have suggested AF ordering with gapless spin-wave excitations.

In contrast to $\chi_{core}$ the term $R_0$ depends on the applied magnetic field.
A natural extension of the latter term to the region outside the cores suggests that it describes relaxation that depends
on thermal quasiparticles through the Zeeman and Doppler effects which are field dependent, but not temperature dependent.  We
note that at our highest field, 42 T, it appears that
$R_0$ decreases, possibly from additional suppression of the superconducting order parameter at high field,
reducing the Doppler term.

It might seem surprising  that $^{17}T_1$ is sensitive to antiferromagnetic spin fluctuations at low temperatures in the
superconducting state since in the normal state a geometric form factor  screens the oxygen nucleus from these
fluctuations\cite{takigawa89,mitrovic02}.  There are two reasons for this\cite{mitrovicthesis}.  At low temperatures the
superconducting energy gap constrains possible  scattering processes  to the nodal regions.  This severely limits the available
phase space and then screening of oxygen by form factors is much less effective.  Secondly, incommensurability of the susceptibility
in the superconducting state\cite{dai01} increases oxygen sensitivity to antiferromagnetic fluctuations. 

Our measurements show that the oxygen NMR relaxation rate, in the form $(^{17}T_1T)^{-1}$, is
enhanced in the vortex core. This indicates the existence in the core of correlated antiferromagnetic fluctuations near a quantum
critical point with a small spin-gap.  Outside the core the relaxation rate is proportional to the temperature at low temperatures,
and can be attributed to the nodal quasiparticles of a {\it d}-wave superconductor. 

We thank  M. Eschrig, A. Chubukov, S. Sachdev and S. C. Zhang for useful discussions. The work was supported in part by the National
High Magnetic Field Labratory and the National Science Foundation DMR95-27035 and the State of Florida.

\vspace{-0.2cm}
\bibliographystyle{unsrt}

\begin{thebibliography}{10}
\bibitem{foot}
It has been suggested that a field-sweep can distort the NMR spectra and current distributions and
thereby affect the relaxation rates\cite{kakuyanagi01}. We have compared field-sweep spectra, obtained as described in Ref.
\cite{mitrovic01} where the sample was cooled to 10 K in zero-field prior to starting the field sweep, 
 to spectra obtained after the sample was annealed at 100 K before cooling back to 10 K at each field step. 
 We found that the zero-field cooled spectra are broader for $H_{int}<0$, but the rates extracted from
spectra for $0< H_{int}<0.032$ T are identical regardless of the data acquisition procedure.
\end{thebibliography}

\vspace{0.3cm}

\end{multicols}
\end{document}